\documentclass[10pt,fleqn,a4paper,twoside]{article}
\usepackage{abcm}
\usepackage{amsmath}
\usepackage{amssymb}
\usepackage{pdfcomment}
\def\shortauthor{A. S. Albino, L. C. S. Jardim, D. C. Knupp,  O. M. Pires, A. J. Silva Neto and E. G. S. Nascimento}
\def\shorttitle{Solving partial differential equations on near-term quantum computers}

\begin{document}
\hspace*{-2.5mm}\begin{tabular}{||p{\textwidth}}
\begin{center}
\vspace{-4mm}
\title{Solving partial differential equations on near-term quantum computers}
\end{center}

\\

\authors{Anton Simen Albino} \\
\institution{SENAI CIMATEC, HPC center - Av. Orlando Gomes, 1845 - Piatã, Salvador - BA} \\
\institution{anton.albino@fbter.org.br} \\

\\
\authors{Lucas Correia da Silva Jardim} \\
\authors{Diego Campos Knupp} \\
\authors{Antônio J. Silva Neto} \\
\institution{UERJ – Instituto Politécnico do Rio de Janeiro – Rua Bonfim, 25, Vila Amélia, Nova Friburgo – RJ} \\ 

\\

\authors{Otto Menegasso Pires} \\
\authors{Erick Giovani Sperandio Nascimento} \\

\institution{SENAI CIMATEC, HPC center - Av. Orlando Gomes, 1845 - Piatã, Salvador - BA} \\
\institution{otto.pires@fbter.org.br, erick.sperandio@fieb.org.br} \\
\\
\abstract{\textbf{Abstract.} In this work, we obtain the numerical temperature field to a thermally developing fluid flow inside parallel plates problem with a quantum computing method. The physical problem deals with the heat transfer of a steady state, hydrodinamically developed and thermally developing fluid flow inside two parallel plates channel subjected to a prescribed constant heat flux. Its solution is formulated numerically with Finite Differences method, where a sequence of linear systems must be solved in order to determine the complete temperature field. Such linear systems are written as discrete unconstrained optimization problems with 
floating points being approximated using binary variables and solved using near-term quantum heuristics. Due to the exponential cost of simulating quantum algorithms, a reduced number of qubits had to be used in the simulations, causing a loss of precision in the results. However, this work advances the state of the art of solutions of differential equations with noisy quantum devices and could be used for useful applications when quantum computers with thousands of qubits become available.}  \\

\\
\keywords{\textbf{Keywords:} Quantum Computing, Quantum Approximate Optimization, Partial Differential Equations, Convection Heat Transfer, Finite Differences }\\
\end{tabular}

\section{INTRODUCTION}

Computational Heat Transfer emerges in the sixties as a way to solve problems that were previously intractable analytically. As the decades progressed, many advances in computational techniques have been made, such as numerical methods for differential equations, increasingly efficient optimization methods, artificial intelligence applied to physics, and many others. All these efforts were allied to the stellar development of the digital computer in order to obtain increasingly faster and more reliable solutions. Nevertheless, a physical limit exist in this scneario, which is the way computers work today. In \cite{limitscomputing}, the physical limits of current computing are explained quantitatively.

Quantum computing was proposed by \cite{feynman} as an alternative for solving intractable computational tasks for classical computing, taking advantage of quanta systems properties such as superposition, entanglement and interference. \cite{deutsch} formalized the notion of a quantum computer, an important advance in quantum computing. Later, \cite{shor}, worked out a very efficient way of performing a Fourier transform using a quantum computer, and applied it to formulate an efficient quantum algorithm for factoring large numbers. According to \cite{preskill}, advances on building quantum devices were made, until reaching the era of Noisy Intermediate Scale Quantum (NISQ) devices. Recently, algorithms capable of dealing with the noise of quantum computers have been developed. Examples of these algorithms and their applications can be seen in the work proposed in \cite{vqa}.

Theoretical approaches to solving nonlinear differential equations in fault-tolerant quantum computers can be seen in (\cite{lloyd, hhl}). Evidence of exponential speed-up in solving linear systems can be seen in \cite{hhl}, which could also be applied to solutions of discretized differential equations. However, the algorithm assumes the possibility of efficiently preparing quantum states and, for its implementation, it is also necessary to decompose matrices into Pauli operators, whose cost is also exponential. In the work of \cite{vqls}, a quantum variational algorithm approach was created for the solution of linear equations, in which state preparations and matrix decomposition in Pauli operators are also required.

The main objective of this work is to use a quantum algorithm in order to solve a two dimensional heat transfer problem that deals with a steady state, hydrodinamically developed and thermally developing fluid flow inside two parallel plates channel subjected to a prescribed constant heat flux. The mathematical formulation includes convection in one direction, where the flow velocity is known. Its solution is formulated numerically with Finite Differences method. Firstly, the problem is described and formulated with all the boundary conditions. This yields a partial differential equation with two spatial dimensions. Secondly, the equation is discretized with derivative approximations, yielding a linear system that must be solved sequentially. The complete solution of these systems yields the two dimensional temperature profile inside the channel. 
In order to obtain the solution of these linear systems with a quantum algorithm, such systems are modeled as unconstrained discrete optimization problems, writing the real variables through floating point approximations, and solved using a heuristic for near-term quantum computers called the Quantum Approximate Optimization Algorithm (QAOA). Our contribution is in the modeling and implementation of quantum algorithms for the solution of any nonlinear partial differential equations in NISQ computers. The amount of variables involved in the floating point approximation of the solution is directly related to the accuracy of the algorithm. The number $n$ of quantum bits needed to implement the model scales linearly with the order $M$ of the system matrices and has a factor $R$ that represents the number of quantum bits used in the floating point approximation. Therefore, we can write the complexity in number of qubits as $n = \mathcal{O}(R \times M)$ which can also be thought of as $n \approx \mathcal{O}(M)$ since the factor $R$ is not scales directly with the number of variables.

\section{PHYSICAL PROBLEM FORMULATION AND SOLUTION}

\begin{figure}[h!]
\centering
\includegraphics[width=80mm]{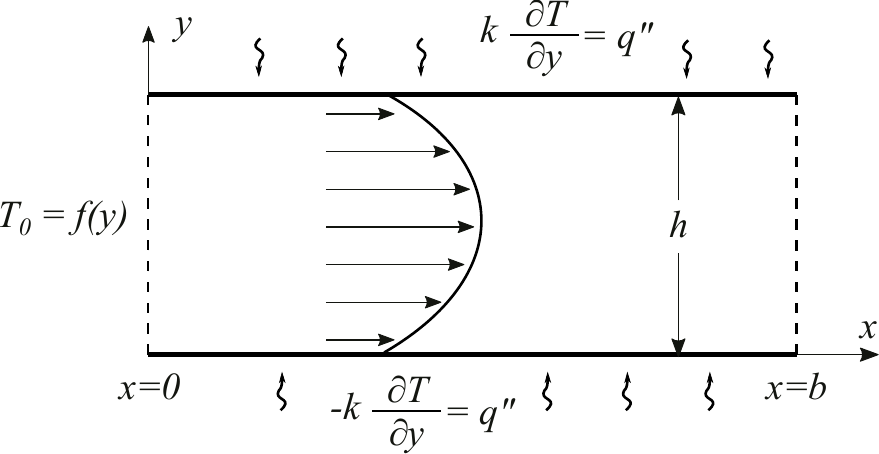}
\caption{Schematic representation of the parallel plates system showing the height dependence of the fluid velocity with the transversal coordinate}
\label{fig:schematic}
\end{figure}

Consider a hydrodinamically developed and thermally developing fluid flow inside a two parallel plates channel subjected to a constant heat flux $q^"$ \citep{bejan2013, shah1978}, as presented schematically in Fig. (\ref{fig:schematic}). The temperature field $T(x, y)$ can be spatially described with the coordinates $x$ and $y$, being the fluid velocity $u(y)$ a function of the vertical coordinate $y$. 

\subsection{Mathematical formulation}

In the formulation of the heat transfer problem, viscous dissipation, free convection, and axial conduction effects are neglected. Furthermore, all the thermal and physical parameters are considered constant. Based upon such hypothesis, the mathematical formulation that describes the temperature field $T(x,y)$ can be written as \citep{bokar}

\begin{subequations}
	\label{eq:model}
	
	\begin{equation}
	k \frac{\partial^2 T (x,y)}{\partial y^2} = u(y) \rho c_p \frac{\partial T (x,y)}{\partial x}, \; \textrm{in} \; 0 < x < b \; \textrm{and} \; 0 < y < h
	\end{equation}
	
	\begin{equation}
	-k\left.\frac{\partial T (x,y)}{\partial y}\right\rvert_{y=0}=q^", \; \textrm{with} \; 0 < x < b
	\end{equation}
	
	\begin{equation}
	k\left.\frac{\partial T (x,y)}{\partial y}\right\rvert_{y=h}=q^", \; \textrm{with} \; 0 < x < b
	\end{equation}
	
	\begin{equation}
	T(0,y) = f(y), \; \textrm{in} \; 0 \leq y \leq h
	\end{equation}
	
	\begin{equation}
	u(y) = 6 u_m \frac{y}{h} \left( 1 - \frac{y}{h} \right)
	\end{equation}
	
\end{subequations}

\noindent where $k$ is the thermal conductivity, $u(y)$ is the fluid velocity field, $\rho$ the specific mass, $c_p$ the specific heat, $f(y)$ is the inlet temperature profile, $h$ is the height of the channel, and $b$ its length. The fluid velocity $u(y)$ is parabolic and the no-slip condition is considered, where, at the boundaries, the fluid assumes zero velocity relative to the boundary.

\subsection{Numerical solution}

\begin{figure}[h!]
\centering
\includegraphics[width=60mm]{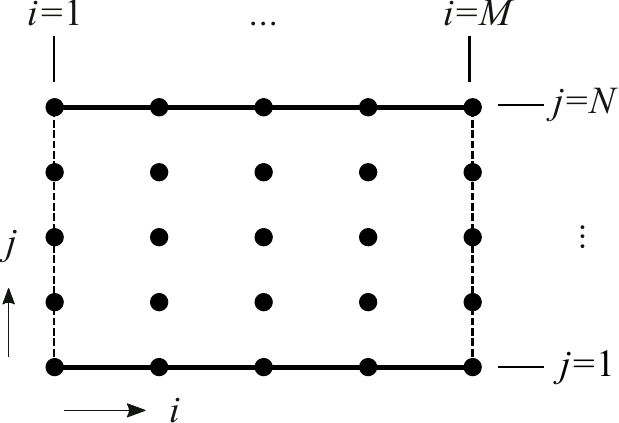}
\caption{Schematic representation of the computational mesh used to discretize the mathematical domain}
\label{fig:mesh}
\end{figure}

To numerically solve the problem described by Eq. (\ref{eq:model}), consider the mesh presented in Fig. (\ref{fig:mesh}), where the nodes $i$ and $j$ are related to spatial coordinates $x$ and $y$, respectively.

Using a finite difference implicit formulation, centered for $y$ and backwards for $x$, the second and first order derivative approximations at nodes $(i+1, j)$ can be written as \citep{ozisik2017}

\begin{subequations}
	\label{eq:finitedifferences}
	
	\begin{equation}
	\frac{\partial^2 T (x,y)}{\partial y^2} \approx \frac{T_{i+1, j+1} - 2 T_{i+1, j} + T_{i+1, j-1}}{\Delta y^2}
	\end{equation}
	
	\begin{equation}
	\frac{\partial T (x,y)}{\partial x} \approx \frac{T_{i+1, j} - T_{i, j}}{\Delta x}
	\end{equation}
	
\end{subequations}
\noindent
where $\Delta x$ and $\Delta y$ are the incremental step sizes in the $x$ and $y$ directions. Replacing the derivative terms in Eq. (\ref{eq:model}) by the approximations presented in Eqs. (\ref{eq:finitedifferences}), the following expression is obtained

\begin{equation}
\label{eq:middle}
	-r_j T_{i+1, j-1} + \left( 2r_j + 1 \right) T_{i+1, j} - r_j T_{i+1, j+1}	= T_{i, j} \; \textrm{for} \; i=2,3,...,M-1 \; \textrm{and} \; j=2,3,...,N-1
\end{equation}
\noindent
where $r$ is a group of parameters that have a dependence in the $y$ coordinate, due to velocity $u(y)$, defined as

\begin{equation}
\label{eq:r}
    r_j = \frac{k}{\rho c_p} \frac{\Delta x}{\Delta y^2} \frac{1}{u_j} \; \textrm{for} \; j=2,3,...,N-1
\end{equation}

Forward and backward differences are used to discretize the first and last boundary at $y=0$ and $y=h$, respectively, and, in order to maintain the same truncation order error given by discretizations with differences presented in Eq. (\ref{eq:finitedifferences}), the following three point differences are used

\begin{subequations}
	\label{eq:boundary_numerical}
	
	\begin{equation}
	\left.\frac{\partial T (x,y)}{\partial y}\right\rvert_{y=h} \approx \frac{T_{i+1, N-2} - 4 T_{i+1, N-1} + 3 T_{i+1, N}}{2 \Delta y}
	\end{equation}
	
	\begin{equation}
	\left.\frac{\partial T (x,y)}{\partial y}\right\rvert_{y=0} \approx \frac{-3 T_{i+1, 1} + 4 T_{i+1, 2} - T_{i+1, 3}}{2 \Delta y}
	\end{equation}
	
\end{subequations}
\noindent
and, replacing them in the boundary conditions of Eq. (\ref{eq:model}), it is possible to obtain

\begin{subequations}
	\label{eq:boundaryfinal}
	
	\begin{equation}
	T^{N-2}_{i+1} - 4 T^{N-1}_{i+1} + 3 T^{N}_{i+1} = \frac{q^"}{k} 2 \Delta y
	\end{equation}
	
	\begin{equation}
	-3 T^{1}_{i+1} + 4 T^{2}_{i+1} - T^{3}_{i+1} = - \frac{q^"}{k} 2 \Delta y
	\end{equation}
	
\end{subequations}
\noindent
and finally, combining Eqs. (\ref{eq:middle}) and (\ref{eq:boundaryfinal}), it is possible to write the following linear system 

\begin{equation}
\label{eq:linearsystem}
\begin{bmatrix}
-3      & 4       & -1      & 0       & \cdots & 0 \\
-r_2      & (2r_2+1)  & -r_2      & 0       & \cdots & 0 \\
0       & -r_3      & (2r_3+1)  & -r_3      & \cdots & 0 \\
0       & \cdots  & \ddots  & \ddots  & \ddots & 0 \\
0       & \cdots  & 0       & -r_{N-1}      & (2r_{N-1}+1) & -r_{N-1}\\
0       & \cdots  & 0       & 1       & -4     & 3 
\end{bmatrix}
\begin{bmatrix}
T_{1}\\T_{2}\\T_{3}\\T_{4}\\ \vdots \\T_{N}
\end{bmatrix}_{i+1}
=\begin{bmatrix}
-q^" 2 \Delta y / k\\T_2\\T_3\\\vdots\\ T_{N-1} \\ q^" 2 \Delta y / k
\end{bmatrix}_i
\end{equation}
\noindent
which yields as solution the vertical temperature points at the ``next" step $i+1$, therefore making it a marching problem that is necessary to solve for $i=1, 2, M-1$.


\section{QUANTUM COMPUTING}

Quantum computing uses principles of quantum mechanics to perform computational tasks. For this to be possible, quantum systems of two discrete energy levels are used (e.g. spins in quantum dots, superoconducting circuits, trapped-ions, etc.) to form a computational basis, $B = \{0.1\} $, where each energy level corresponds to one of the base states. In quantum computing, the basic unit of information is the quantum bit, or qubit, defined mathematically as

 \begin{equation}\label{qubit}
    |\psi\rangle = \alpha_0|0\rangle + \alpha_1|1\rangle = \begin{bmatrix}
    \alpha_0 \\
    \alpha_1
    \end{bmatrix}
\end{equation}   
where $|0\rangle$ and $|1\rangle$ are vectors represented in Dirac's notation and $\alpha_0$ and $\alpha_1$ are complex numbers that must satisfy the normalization condition $|\alpha_0|^2 + | \alpha_1|^2 = 1$. The qubit state conjugate transposed vector is represented in Dirac notation as $\langle \psi | = (\alpha_0^* \ \ \alpha_1^*)$. Such vectors form an orthonormal basis $b = \{|0\rangle, |1\rangle\}$. The vectors of the computational basis can be represented in the standard form of linear algebra as

$$ |0\rangle = \begin{bmatrix}
                 1\\
                 0
     \end{bmatrix} \ \   \text{and} \ \  |1\rangle = \begin{bmatrix}
                 0\\
                 1
     \end{bmatrix}. $$
     
In systems involving multiple qubits, we represent the overall state of the system as a tensor product of the state of each of the $n$ qubits, such as $|\psi_0\psi_1...\psi_n\rangle \equiv |\psi_0\rangle \otimes |\psi_1\rangle \otimes ...\otimes |\psi_n\rangle$. The quantum state involving multiple qubits can also be written more compactly, in decimal base, as

\begin{equation}
    |\psi_0\psi_1...\psi_n\rangle = \sum_{i=0}^{2^n-1}\alpha_i|i\rangle,
\end{equation}
which as can be seen, can be written as a linear combination of $2^n$ eigenstates of the computational base. This means that it is a high-dimensional space, the Hilbert space, which scales exponentially with the number of qubits. The physical explanation of the linear combination of these vectors is due to the superposition of states, which makes it possible to create and operate on all eigenstates simultaneously. An operation, $U$, on an initial quantum state, $|s \rangle$, produces a linear transformation such that

\begin{equation}
    U|s\rangle =  |\psi_f\rangle.
\end{equation}
The mathematical representation of quantum operators is given through unitary matrices ($U^\dagger U = I$). Therefore, a quantum circuit is a successive set of unitary operations on any initial state. A quantum operator acting on a superposition state will be acting on all eigenstates simultaneously. A quantum operator can also be factored as a tensor product of other lower-order operators, such as $U = U_0\otimes U_1 \otimes ... \otimes U_n$. In a system with $n$ qubits, an operation involving all of them is represented by a unit matrix of order $2^n \times 2^n$.

In quantum computing, a measurement on a quantum state is probabilistic and can be represented as a projection of the measured vector on the computational basis. Each eigenstate of the computational basis, $|i\rangle$, of a quantum state, $|\phi\rangle$, has a probability of being measured, which can be calculated by Born's Rule, as
\begin{equation}
    p(|i\rangle) = |\langle i|\phi\rangle |^2 = |\alpha_i|^2.
\end{equation}
Therefore, to characterize a quantum state in terms of probability amplitudes, a sufficient set of measurements must be performed on the state. However, quantum algorithms must be developed in such a way that a relatively small number of measurements are necessary. A review on basic quantum computing and algorithms can be seen in \cite{portugal}.

\subsection{Variational Quantum Algorithms}

Quantum algorithms have been proposed to investigate potential advantages in optimization problems. Due to the need for fault-tolerant and error-corrected quantum computers for the implementation of some quantum algorithms, in this work, we will keep our attention on algorithms capable of being implemented in Noisy Intermediate Scale Quantum (NISQ) devices, which can be applied in near-term.The main idea behind quantum approximate optimization is to optimize a quantum circuit to obtain the lowest expected energy value of a given Hamiltonian cost, $\mathcal{H}$, written as

\begin{figure}[h!]
\centering
\includegraphics[width=120mm]{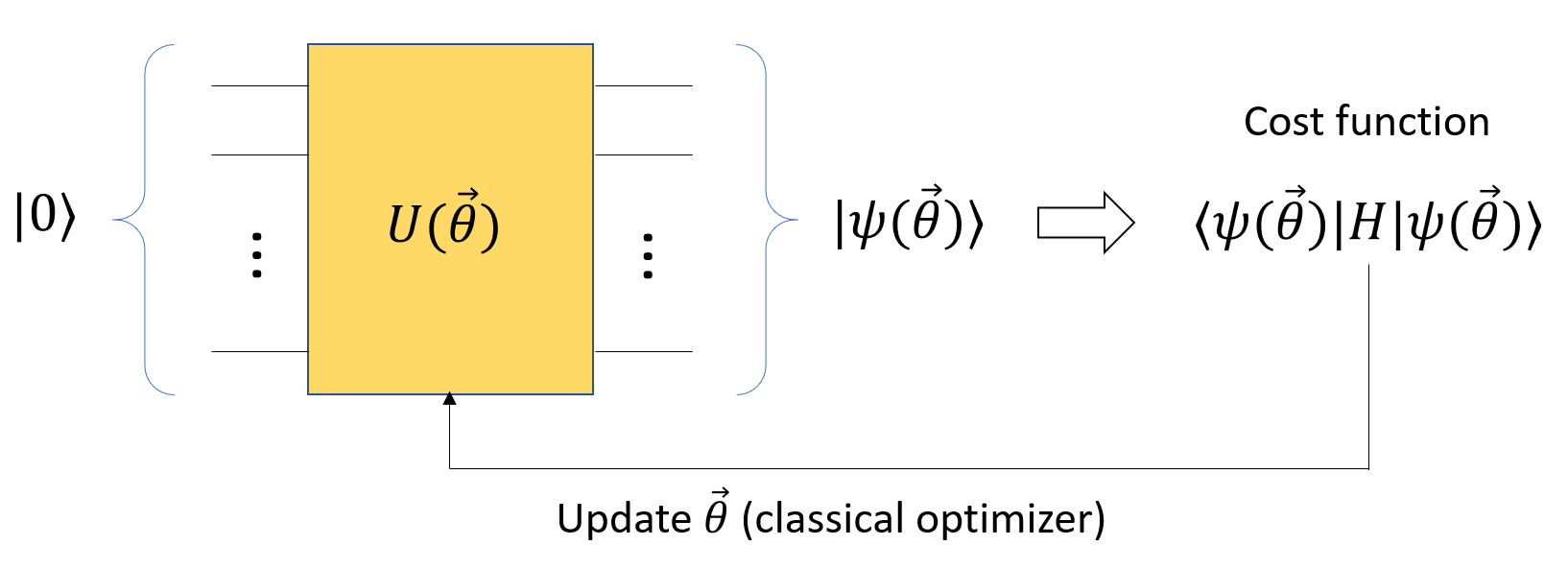}
\caption{Schematic workflow representing Variational Quantum Algorithms.}
\label{fig:vqa}
\end{figure}

\begin{center}
 \begin{equation}
    \langle \psi | \mathcal{H} | \psi \rangle  = \sum_{x=0}^{}E_ip_{|i\rangle},
\end{equation}   
\end{center}
where $E_i$ and $p_{|i\rangle}$ are the energy associated to the eigenstate $|i\rangle$ and its probability, respectively. Such algorithms are mostly Variational Quantum Algorithms (VQAs), which can be seen in \cite{vqa, vqe, qaoa}, where a quantum circuit, parametrized by rotation angles, $\Vec{\theta}$, is created in order to generate a statevector $|\psi(\Vec{\theta})\rangle$. The parametrized quantum circuit is oftenly called ansatz, which can be implemented as a heuristic pattern using rotation single quantum gates and controlled-NOT operators. Angles from rotation gates are updated iteratively through classical optimization untill the minimum expected value is obtained as can be seen in Fig. \ref{fig:vqa}. 

The cost hamiltonian, $H$, is an objective function written as an Ising Model, whose spin variables are $z_i \in \{-1,+1\}$. Therefore, it is an unconstrained optimization problem with binary discrete variables. A Quadratic Unconstrained Binary Optimization (QUBO) can be mapped to an Ising Model and vice-versa through the relation $z_i = 2s_i - 1$, where $s_i \in \{0,1\}$ are QUBO variables. 

\subsection{Quantum Approximate Optimization Algorithm (QAOA)}

The Quantum Approximate Optimization is a variational algorithm of quantum computing, developed by \cite{qaoa}, suitable for solving discrete optimization problems. The algorithm consists of modeling the objective function as an \textit{Ising Hamiltonian} and transforming it into unitary operators, applying such operators in an input state, measuring the expected values, updating the parameters and repeating the process until that the cost function is minimized. When the energy of the objective function is minimized, the probability distribution (obtained after measuring the wavefunction $|\psi\rangle$ with the optimal parameters) has a higher occurrence of the eigenstate of $|\psi\rangle$ which gives the best configuration for the problem.

\subsubsection{Ising model}

The model of the behavior of ferromagnetic materials can be described by the \textit{Ising Hamiltonian}. The model represents a chain of spin 1/2 particles, which interact in pairs, each constituent being subject to a magnetic field of magnitude $h_i$. As spin 1/2 particles can assume only two dipole moment directions ("up" and "down"), the interaction between each pair of particles provides an energy (positive or negative) according to the spin orientations. To find the configuration of the dipole moments that provides the lowest possible energy value of the system, the following objective function must be minimized: \newline
\begin{equation} \label{fo_ising}
    C(z) = - \sum_{<ij>}^{}z_iz_j - \sum_{i}^{}h_iz_i .
\end{equation} \newline
The variables $z_i$ of the objective function can only take on unit values, whether positive or negative, that is, $z_i \in \{-1,+1\}$. The variables can be written as Pauli-Z operators ($\sigma_z$), since $\pm 1$ are eigenvalues of $\sigma_z$, associated with the spin eigenfunctions of the particles.

\subsubsection{Gamma Operator}

The Gamma Operator (or Phase Operator) represents the interactions between adjacent particles, as well as the response of the particles to the applied magnetic field. It is about writing the objective function of the problem (Equation \ref{fo_ising}) as a unit operator so that it can be applied directly in the circuit. As seen, the operator described in Equation \ref{fo_ising} is non-unitary, since it does not satisfy the criterion of Equation 7. However, each term of the sum (neglecting the coefficients) is a unitary operator. To describe this Hamiltonian, the Gamma Operator is defined through the relation
\begin{equation} \label{3.14}
    U(C,\gamma)=e^{-i\gamma C(z)}=e^{i\gamma\sum_{<ij>}^{}z_iz_j + i\gamma\sum_{i}^{} h_iz_i} .
\end{equation}
As $C(z)$ is expressed in terms of the Pauli-Z operators and the tensor product between them generates diagonal matrices, $U(C,\gamma)$ is also a diagonal matrix (but unitary), where the parameter $\ gamma$ is one of the variational parameters used in QAOA. Thus, the operator $U(C,\gamma)$ can be written as the unit operation
\begin{equation} \label{3.15}
    U(C,\gamma)=\prod_{<ij>}^{}e^{i\gamma Z_iZ_j}\prod_{i=0}^{N}e^{i\gamma h_iZ_i} .
\end{equation}
This operator (as well as tensor products between Pauli-Z matrices) takes the form of a diagonal matrix, where the diagonal elements are complex exponentials that add phases in the quantum state in which it is applied.
\subsubsection{Mix Operator}
The operator $U(B,\beta)$ is applied with the objective of mixing the amplitudes (generating destructive and constructive interferences) making it possible to better explore the search space and increasing the chances of finding, with greater precision, the global optimal point of the objective function. The Mixing Operator can be described mathematically as
\begin{equation} \label{mix}
    U(B,\beta) = e^{-i\beta \sum_{j=0}^{N} X_j} \equiv \prod_{j=0}^{N}e^{-i\beta X_j} .
\end{equation}
Such an operator can be constructed using only rotation operations on individual \textit{qubits} of type RX. However, the efficiency of this operator is directly linked to the convergence time and the accuracy of the algorithm. The \ref{mix} equation describes the operator in its simplest form, represented only as rotations around the $x$ axis of the \textit{Bloch Sphere}.

\subsubsection{QAOA objective function: Measuring Expected Values}
The unit operations presented in the previous subsections can be represented as quantum circuit diagrams. In QAOA, each set of the operations in question are known as \textit{layers}. The $i$th \textit{layer} of the circuit can be written as
$
    U(\gamma_i,\beta_i) = U(B,\beta_i)U(C,\gamma_i),
$
and, therefore, the sequence of these operators represents the general operator of the QAOA. The greater the number of \textit{layers}, the greater the number of circuit parameters and, consequently, the greater the accuracy of the result. As with VQE, the objective function of QAOA is also an expected value. The difference between them is that, in the VQE, a generic ansatz is used and in the QAOA, in turn, the ansatz is constructed from the objective function. The performance of the QAOA general operator on the input state $|s\rangle$ can be considered as follows:

\begin{equation} \label{cost_qaoa}
    |\vec{\gamma},\vec{\beta} \rangle = (U(B,\beta_n)U(C,\gamma_n))...(U(B,\beta_1)U(C,\gamma_1 ))(U(B,\beta_0)U(C,\gamma_0))|s\rangle.
\end{equation}

As it is of interest to find the lowest energy configuration for the Hamiltonian, the objective function of the problem is, again, the function of the expected energy values, which, if minimized, returns the lowest energy eigenvalue of the system. The objective function of the problem can be expressed as

\begin{equation} \label{obj-func}
    \langle C(Z) \rangle \equiv \langle\vec{\gamma},\vec{\beta}|C(Z)|\vec{\gamma},\vec{\beta} \rangle.
\end{equation}

As seen in previous chapters, the expected value of energy can also be written as an average value. When the quantum state is prepared and measured a sufficient number of times, the probability distribution obtained allows us to calculate the expected value as the sum of the number of times each eigenstate of the wavefunction was measured, multiplied by their respective associated energies. The variational parameters, $\gamma$ and $\beta$, are updated iteratively through classical optimization methods until the lowest energy eigenvalue is found. As these are combinatorial optimization problems, the problem is concerned with the eigenstate of the computational base associated with the smallest eigenvalue, which is composed of the \textit{bitstring} formed by the optimal configuration of the problem. As seen, the QAOA can be used to find the configuration that minimizes the energy of the \textit{Ising Hamiltonian}. As it is an optimization problem with discrete variables, it is clear that the algorithm is a good candidate to solve problems of this nature.

\subsection{Warm-start quantum approximation}
    One of the biggest challenges for variational quantum algorithms to overcome is the barren plateau phenomenon, in which the cost for the parameter training becomes prohibitively high as the number of parameters increase due to the vanishing of the cost function partial derivatives~\cite{Cerezo2020}.
    One of the main approaches to mitigate the impact of barren plateaus is to choose a good parameter initialization strategy. Poor initialization strategies, as randomly initializing the parameter values, can lead to the algorithm starting far from the solution, near a local minima. More refined initialization strategies were proposed which were shown to have an exponentially advantage over random initialization~\cite{Zhou2020}.
    
    Warm-starting quantum optimization developed by~\cite{Egger_2021} is a strategy applied to integer programming and combinatorial optimization problems. First the problem to be optimized is relaxed, by replacing binary variables with continuous ones, and a classical solver is utilized to solve the problem in linear time. After that, the solution for the relaxed problem is used as the basis for the first set of parameters in the VQA. This strategy allows the quantum algorithm to inherit the performance guarantees of the classical algorithm.

\section{LINEAR SYSTEM PROBLEM AS AN ISING MODEL}
The quantum approximate optimization described in the previous section is suitable for unconstrained optimization problems. Since to solve the two dimensional convection heat transfer problem, discretized in finite elements, we need to solve successive linear systems of the type $A\Vec{s}=\Vec{b}$, the initial strategy is to transform the problem of solving linear systems into an unconstrained optimization problem, as an objective function written as
\begin{equation}\label{lsp}
    f(\Vec{s}) = \|A\Vec{s} - \Vec{b}\|^2.
\end{equation}
Here, the variables are such that $\Vec{s} \in \mathbb{R}$, however, to deal with this evolutionary heuristic, we are interested in dealing with binary variables. Therefore, we will use an R-bit approximation to write the components of the vector $\Vec{s}$ as an approximation using binary variables. As the solution vector is composed of temperatures in Kelvin (K) of points on the mesh, we can only worry about approximating positive real numbers. In this case, each variable $s_i$ can be written in terms of binary variables $q_{i,r}$ through the R-bit approximation as $s_i = \sum_{r=-R}^{R}q_{i,r}2^{r}$. However, an Ising model cost function is required to mapping such cost function into unitary quantum operators easily. Thus, the (Eq. \ref{lsp}) can be rewrite the R-bit approximation in terms of spin varibles as  $s_i = \sum_{r=-R}^{R}(\frac{z_{i,r}+1}{2})2^{r}$ and the objective function is written following discrete optimization problem
\begin{equation}\label{qubo-lsp}
    q^*  = \underset{q \in \{0,1\}^{R\times M}}{\operatorname{arg \ min}} \ \sum_{k=1}^{M}\left(\sum_{i=1}^{M}a_{k,i}\left(\sum_{r=-R}^{R}\left(\frac{z_{i,r}+1}{2}\right)2^{-r}\right) - b_k\right)^2
\end{equation}
whose varibles are $z \in \{-1,+1\}$. Therefore, the problem of solving a linear system has been mapped to an Ising model or QUBO. A quantum eigenstate, written in the computational basis can be evaluated in the objective function (Eq. \ref{qubo-lsp}) and the result is interpreted as the energy associated with the given eigenstate. The mean value of the all measured eigenstates is the expected value of energy (Eq. \ref{obj-func}).


\section{RESULTS AND DISCUSSION}

To obtain the results presented in this work for both the regular and quantum computing approaches, all the simulations consider the fluid as air, with parameters and conditions as presented in Tab. \ref{tab:parameters} \citep{liu}. The increments $\Delta x$ and $\Delta y$ are chosen in order to obtain the total number of $N = 5$ and $M=5$, which generates a discretized mesh of $5 \times 5 = 25$ points. The inlet temperature $T(0,y) = f(y)$ is defined as a step function with values 4 ºC for $0\leq y \leq h/2$ and 8 ºC for $h/2 < y \leq h$.

\begin{table}[!h]
\centering
\caption{Thermal and physical parameters used to solve the heat trasnfer problem.}
\begin{tabular}{cccl}
\hline
Parameter  & Value  & Dimension      & \multicolumn{1}{c}{Observation} \\ \hline
$k$        & 0.0265 & {[}W/m.K{]}    & Thermal conductivity            \\
$c_p$      & 2000.0 & {[}J/kgK{]}    & Specific heat                   \\
$\rho$     & 1.1614 & {[}kg/m$^3${]} & Specific mass                   \\
$h$        & 0.01   & {[}m{]}        & Spacing between plates          \\
$b$        & 1.0    & {[}m{]}        & Plate length                    \\
$u_m$      & 2.0    & {[}m/s{]}      & Fluid velocity                  \\
$q^"$      & 50.0   & {[}W/m{]}      & Heat flux at the plates         \\
$\Delta x$ & 20     & {[}m{]}        & Spatial increment in $x$        \\
$\Delta y$ & 0.50   & {[}m{]}        & Spatial increment in $y$        \\ \hline
\end{tabular}
\label{tab:parameters}
\end{table}

\begin{figure}[h!]
\centering
\includegraphics[width=160mm]{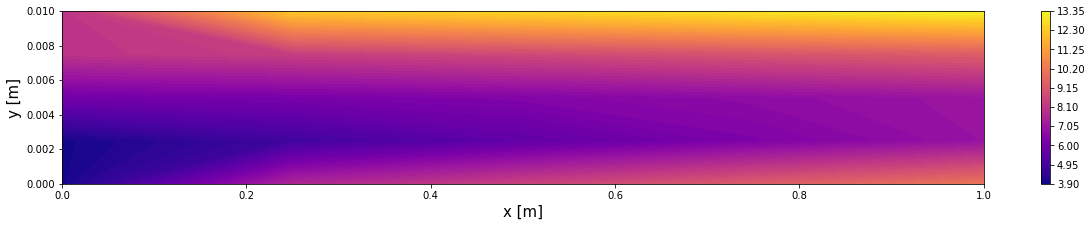}
\caption{Two-dimensional temperature field simulated through numerical methods of classical computation. The vertical and horizontal axes vary, respectively, as $0.00 \leq y \leq h$ and $0.0 \leq x \leq b$.}
\label{fig:tempfieldclassical}
\end{figure}

\begin{figure}[h!]
\centering
\includegraphics[width=160mm]{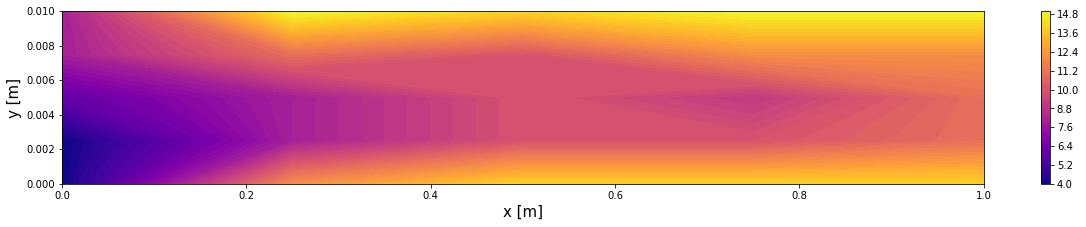}
\caption{Two-dimensional temperature field simulated using QAOA. In this simulation, the floating point approximation was performed using 5 qubits (R=5) only for integers and therefore, $r \in \{0,1,2,3,4\}$, limiting the precision of the solution.}
\label{fig:tempfieldquantum}
\end{figure}

In this experiment, an iterative routine for solving linear systems was implemented warming-start quantum optimization. The experiment was carried out using 5-bit approximation with $r = \left[0,4\right]$ (only integer representation of the temperatures) in order to saving computational resources, since simulating quantum circuits has exponential computational cost associated. The two dimensional heat transfer problem was discretized in order to create $5 \times 5$ matrices. The required number of qubits, in that settings, are $R\times M = 20$. In order to compare the solution found by both (classical and quantum) algorithms, the 2D temperature profiles obtained can be seen in Fig. \ref{fig:tempfieldclassical} and Fig. \ref{fig:tempfieldquantum}.

The choice to simulate using a $5-bit$ floating point approximation is due to the exponential cost of simulating quantum computing. Each operation in a quantum circuit with $n$ qubits acts on a state vector of dimension $2^n$. However, a similar behavior was observed in the temperature field, although the error has propagated since the first iteration, since the approximation was only by integers.

\section{CONCLUSIONS}

Quantum computing has the potential to offer an advantage over classical computing due to quantum properties that make it possible to create and manipulate higher-order states. However, to develop algorithms that take advantage of these phenomena, it is necessary to deal with a new computing paradigm. In this work, we discuss a way to model nonlinear partial differential equations to be solved with NISQ devices, through the finite difference method and with the use of variational algorithms, which contributes to the advancement of the state of the art in the development and applications of quantum algorithms , since a variety of problems can be solved using the same framework. The results showed that QAOA is able to find satisfactory results in solving linear equations problems, but in this work, because of computational limitations and exponential cost to simulate QAOA, we restricted ourselves to working with a reduced number of qubits, leading to inaccurate results. Although such a lack of precision occurred, this was in fact the expected result. Thousands of qubits will be needed before QAOA and its variants can be used to solve these problems, due to the linear relationship between the dimensionality of the problem and the number of qubits. However, the qubits used will not necessarily be error-corrected due to the characteristics of the heuristic itself, which requires low-depth circuits and few measurements of the final state.

\section{ACKNOWLEDGEMENTS}
Acknowledgements to the Supercomputing Center for Industrial Innovation (CS2i), the Reference Center for Artificial Intelligence (CRIA), and the Latin American Quantum Computing Center (LAQCC), all from SENAI CIMATEC. The authors also acknowledge the financial support provided by FAPERJ - Carlos Chaga Filho Foundation for Research Support of the State of Rio de Janeiro, CNPq - National Council for Scientific and Technological Development, and CAPES - Coordination for the Improvement of Higher Education Personell (Finance Code 001)

\section{REFERENCES} 

\bibliographystyle{abcm}
\renewcommand{\refname}{}
\bibliography{bibfile}

\section{RESPONSIBILITY NOTICE}

The authors are solely responsible for the printed material included in this paper.

\end{document}